\documentstyle[12pt]{article}
\def\be{\begin{equation}}
\def\ee{\end{equation}}
\def\bea{\begin{eqnarray}}
\def\eea{\end{eqnarray}}

\begin{document}

\begin{flushright}
IPM/P-2001/014\\
\end{flushright}

\begin{center}
{\Large{\bf Noncommutativity and the Motion of D$_p$-brane along Itself}}

\vskip .5cm
{\large Davoud Kamani}
\vskip .1cm
 {\it Institute for Studies in Theoretical Physics and
Mathematics (IPM)
\\  P.O.Box: 19395-5531, Tehran, Iran}\\
{\sl e-mail: kamani@theory.ipm.ac.ir}
\\
\end{center}

\begin{abstract}

We consider open strings attached to a moving D$_p$-brane 
with motion along itself,                                                                       
in the presence of the backgrounds $B_{\mu \nu}$-field and a $U(1)$ gauge
field $A_{\alpha}$. The effects of the motion of the brane on the open
string propagator and on the open string variables are studied.
We observe that some free parameters appear in the open string variables
and in the propagator of it.
\end{abstract}
$\;\;\;\;\;\;\;\;\;$PACS: 11.25.-w
\vskip .5cm

There have been attempts to explain the noncommutativity on D-brane
worldvolume through the study of open strings in the presence of background 
fields \cite{1,2}. It is known that Lorentz boosts act on background 
fields. This affects the noncommutativity parameter and the effective
metric of open string. According to these facts, some effects of Lorentz
boosts on the noncommutative string theory are studied \cite{3}.

Instead of Lorentz boosts, one can use the $\sigma$-model action of string 
with velocity terms \cite{4,5,6}. These terms show the velocity of 
the brane that the ends of open string are on it. Some applications of this
method are given in the Ref.\cite{5,7}. In this method, the 
velocity of the brane introduces spontaneously to all quantities  
extracted from the string action, for example the noncommutativity
parameter, the open string metric,
the boundary conditions of the open string and its 
propagator. In this note we consider the motion of the brane along itself.
For example for membrane we observe that
some arbitrary parameters appear in the open
string variables and its propagator.
As expected, for zero speed our results reduce to the known cases.

$boundary\;conditions\;of\;open\;string$

We begin with a $\sigma$-model action for string that 
contains $B_{\mu \nu}$ field and two boundary terms corresponding to 
the gauge field \cite{4} and the velocity of the brane \cite{5,6}
\bea
S &=& \frac{1}{4\pi \alpha'} \int_{\Sigma} d^2 \sigma (g_{\mu \nu}\partial_a
X^\mu \partial^a X^\nu -2 \pi i \alpha' \epsilon^{ab} B_{\mu \nu}
\partial_a X^\mu \partial_b X^\nu )
\nonumber\\
&~& +\frac{1}{2\pi \alpha'}\int_{\partial \Sigma}d \tau (-2\pi i\alpha' 
A_{\alpha}\partial_{\tau} X^\alpha + V_\alpha X^0 \partial_\sigma X^\alpha)\;,
\eea
where $\Sigma$ is the worldsheet of the open string and $\partial \Sigma$
is the boundary of it that is at $\sigma =0$.
Here the open string worldsheet $\Sigma$ has the Euclidean metric 
$\delta^{ab}$. $A_\alpha$ is $U(1)$ gauge field that lives in the 
worldvolume of the brane
and $V^\alpha$ is the velocity of the brane. The set $\{ X^\alpha \}$
specifies the directions of the worldvolume of the brane. Also let the set
$\{ X^i \}$ show the directions perpendicular to the brane.

Assume that the background fields $g_{\mu \nu}$ and $B_{\mu \nu}$ to be 
constant, with the zero elements $g_{\alpha i}=B_{\alpha i} = B_{ij} =0$.
Furthermore let the field strength of the gauge field $A_\alpha$ be 
constant. Vanishing the variation of the action with respect to 
$X^\mu(\sigma , \tau)$ gives the equation of motion and the boundary  
conditions of the string
\bea
&~&(\partial^2_\tau +\partial^2_\sigma) X^\mu(\sigma , \tau)=0\;,\\
&~&\bigg{(} g_{0 \beta} \partial_{\sigma} X^\beta + g_{{\bar \alpha}\beta}
V^\beta \partial_\sigma X^{\bar \alpha} + 2\pi i \alpha'{\cal{F}}_{0 \beta} 
\partial_\tau X^\beta \bigg{)}_{\sigma=0}=0\;,\\
&~&\bigg{(} g_{{\bar \alpha} \beta} \partial_{\sigma} ( X^\beta 
- V^\beta X^0 ) + 2\pi i \alpha'{\cal{F}}_{{\bar \alpha} \beta} \partial_\tau
X^\beta \bigg{)}_{\sigma=0}=0\;,\\
&~& (\delta X^i)_{\sigma=0}=0\;,
\eea
where ${\bar \alpha}$ refers to the spatial directions of the brane, i.e.
${\bar \alpha} \neq 0$, and
\bea
{\cal{F}}_{\alpha \beta} = \partial_\alpha A_\beta - \partial_\beta 
A_\alpha + B_{\alpha \beta}\;,
\eea
is total field strength.

Since the transverse coordinates $\{X^i\}$ can be treated trivially,
we concentrate only on the brane directions.
The boundary conditions (3) and (4) can be written as the following
\bea
(S_{\alpha \beta} \partial_\sigma X^\beta + 2\pi i \alpha' {\cal{F}}_{\alpha 
\beta}\partial_\tau X^\beta )_{\sigma=0}=0\;,
\eea
where the matrix $S$ is
\bea
S_{\alpha \beta} = g_{\alpha \beta} + \omega_{\alpha \beta}\;,
\eea
and the antisymmetric matrix $\omega_{\alpha \beta}$ is 
\bea
&~& \omega_{0 {\bar \alpha}} = -\omega_{{\bar \alpha}0}= 
g_{{\bar \alpha}\beta} V^\beta\;,
\nonumber\\
&~& \omega_{{\bar \alpha}{\bar \beta}} = 0\;.
\eea
Now we use the equation of motion (2) and the boundary condition (7), to
obtain the open string propagator.

$open\;string\;variables\;and\;propagator$

The propagator for the open string can be obtained by solving the equations
of motion
\bea
\partial {\bar \partial}{\cal{G}}^{\alpha \beta}(z, z') = 
- \frac{\pi}{2}\alpha' G^{\alpha \beta} \delta^{(2)}(z-z')\;,
\eea
where $z=\tau+i\sigma$, and $G^{\alpha \beta}$ is the open string metric.
Propagator also should satisfy the boundary conditions
\bea
\bigg{(} (\partial -{\bar \partial}){\cal{G}}^{\alpha \beta}(z,z') + 
2\pi \alpha'(S^{-1}{\cal{F}})^\alpha _{\;\;\;\gamma}(\partial+{\bar \partial})
{\cal{G}}^{\gamma \beta}(z,z') \bigg{)}_{z={\bar z}} = 0\;.
\eea
The solution of the equations of the propagator can be written as
\bea
{\cal{G}}^{\alpha \beta}(z,z') &=& -\alpha' \bigg{[} \frac{1}{2} 
Q^{\alpha \beta} \ln(z-z')+ \frac{1}{2} Q^{\alpha \beta}
\ln({\bar z}-{\bar z'})
\nonumber\\
&~& +\bigg{(}-\frac{1}{2}Q^{\alpha \beta} + G^{\alpha \beta} +
\frac{\theta^{\alpha \beta}}{2\pi \alpha'}\bigg{)} \ln(z-{\bar z'})
\nonumber\\
&~& +\bigg{(}-\frac{1}{2}Q^{\alpha \beta} + G^{\alpha \beta} 
-\frac{\theta^{\alpha \beta}}{2\pi \alpha'}\bigg{)} \ln({\bar z}- z')
-\frac{i}{2\alpha'} \theta^{\alpha \beta} \bigg{]}\;.
\eea
The equations (11) and (12) give the open string metric and the 
noncommutativity parameter as the following
\bea
&~& G^{\alpha \beta} = \bigg{(}(S+2 \pi \alpha' {\cal{F}})^{-1}S 
(S-2 \pi \alpha' {\cal{F}})^{-1} SQ \bigg{)}^{\alpha \beta}\;,\\
&~& G_{\alpha \beta} = \bigg{(}Q^{-1}S^{-1}(S-2 \pi \alpha' {\cal{F}}) S^{-1} 
(S+2 \pi \alpha' {\cal{F}})\bigg{)}_{\alpha \beta}\;,\\
&~& \theta^{\alpha \beta} = -(2\pi \alpha')^2 \bigg{(}(S+2 \pi \alpha' 
{\cal{F}})^{-1}{\cal{F}}(S-2 \pi \alpha' {\cal{F}})^{-1} SQ 
\bigg{)}^{\alpha \beta}\;.
\eea
The open string metric $G$ should be symmetric and the
noncommutativity parameter $\theta$ should be antisymmetric.
These imply that the matrix $Q$ should satisfy the following equation
\bea
SQ{\cal{F}}-{\cal{F}}(SQ)^T = \frac{1}{2\pi \alpha'} S(Q^T-Q)S^T\;.
\eea
The propagator (12) under the exchanges $\alpha \leftrightarrow \beta$ and
$z \leftrightarrow z'$, should be symmetric. Therefore the matrix $Q$ is
symmetric, i.e.
\bea
Q^T = Q\;.
\eea

Since the equations (16) and (17) do not give a unique solution for the
matrix $Q$, some arbitrary parameters appear in the open string variables 
and the propagator of it. In fact these 
free parameters show a class of solutions.

According to the condition (17), $Q=S^{-1}$ 
is not an available solution of the 
equation (16), because $S$ is not a symmetric matrix. For zero speed we have
$Q=g^{-1}$. In this case the propagator (12) and the equations (13)-(15)
reduce to the known cases.
                                                   
Since the ordinary DBI action and the noncommutative description of it at 
zero speed and zero field strength are equal the effective open string 
coupling $G_s$ is as previous, i.e. it is independent of the speed of the
brane.

By using the equations (13)-(17) and the identity
\bea
\partial {\bar \partial} (\ln|z-z'| + \ln|z-{\bar z'|})=\frac{\pi}{2}
\delta^{(2)}(z-z')\;,
\eea
one can verify that the propagator (12) also satisfies the equation (10).

$\alpha' \rightarrow 0\;limit$

The zero slope limit ($\alpha' \rightarrow 0$) of this open string system
depends on the speed of the brane. Since in this limit there are $\alpha'
\sim \epsilon^{\frac{1}{2}}$ and $g_{\alpha \beta} \sim \epsilon$ (where
$\epsilon \rightarrow 0$), the term $\omega_{\alpha \beta}$ in
the equation (8) is zero (for zero elements) or goes to zero like
$\epsilon$ (for non-zero elements). Therefore in this limit
there is $S_{\alpha \beta} \sim \epsilon$. 
From the equations (13)-(15) we have  
\bea
&~& G^{\alpha \beta} = -\frac{1}{(2 \pi \alpha')^2}({\cal{F}}^{-1}S 
{\cal{F}}^{-1} SQ )^{\alpha \beta}\;,\\
&~& G_{\alpha \beta} = -(2 \pi \alpha')^2 (Q^{-1}S^{-1}{\cal{F}} S^{-1} 
{\cal{F}})_{\alpha \beta}\;,\\
&~& \theta^{\alpha \beta} = ({\cal{F}}^{-1}SQ )^{\alpha \beta}\;.
\eea
From the equations (16) and (17) we see that the matrix $SQ$ in this limit
is finite. Therefore the open string metric $G$ and the noncommutativity
parameter $\theta$ also are finite. 

When the speed of the brane is non-zero, the equations (16) and (17) impose 
some conditions on the matrices $S$ and ${\cal{F}}$. 
To see this clearly, we study the following examples. Since for the
Minkowski spacetime, again the equations (16) and (17) hold, we consider the 
following examples in the Minkowski spacetime.

$Example\;1:\;D_1-brane$

For D$_1$-brane with speed
$v$ along itself, $Q$ is non-zero if there is 
$v^2 \det S = 0$. Zero speed $v=0$, is a
known case. The case $\det S =0$, means that the matrix $S$ is not
invertible, and therefore the open string metric (13) is not invertible.
In this case the speed of the D$_1$-brane is
\bea
v = \pm \sqrt{g'^2 + g_0 g_1}\;,
\eea
where the matrix 
$\left( \begin{array}{cc} 
-g_0 & g' \\ 
g' & g_1 
\end{array} \right)$
is the closed string metric. Since $-1 \leq v \leq 1$, the speed (22) is
available if the elements of the closed string metric obey the condition
$g_0 g_1 + g'^2 \leq 1$.

$Example\;2:\;D_2-brane$

Consider a D$_2$-brane parallel to the $X^1X^2$-plane, with the speed $v$
along $X^1$-direction. Let the closed string metric be $diag(-g_0,g_1,g_2)$.
The matrix $Q$ is
\bea
Q=\left( \begin{array}{ccc} 
q & q_1 & q_2 \\ 
q_1 & q' & q_3 \\ 
q_2 & q_3 & q'' 
\end{array} \right)\;.
\eea
If there is an electric field on the brane parallel to the velocity, the
element $q''$ is arbitrary. The other elements of $Q$ are zero.
Therefore ${\cal{G}}^{\alpha\beta}$, 
$G^{\alpha\beta}$ and $\theta^{\alpha\beta}$ 
depend on the arbitrary parameter $q''$. 

If the electric field is 
perpendicular to the velocity, the element $q'$ is
arbitrary and $q_2 = q_3 =0$. Also there are the following relations between
$q_1$, $q$ and $q''$
\bea
&~& q_1 = qv\;,
\nonumber\\
&~& q'' = \frac{-g_0 + g_1 v^2}{g_2}q\;.
\eea
In this case the free parameters $q$ and $q'$ appear on the 
variables ${\cal{G}}^{\alpha\beta}$, 
$G^{\alpha\beta}$ and $\theta^{\alpha\beta}$.

For magnetic field perpendicular to the brane, $q$ is arbitrary and
\bea
&~& q_2 = q_3=0\;,
\nonumber\\
&~& q_1 = \frac{g_1}{g_0}vq'\;,
\nonumber\\
&~& q''=(1- \frac{g_1}{g_0}v^2)\frac{g_1}{g_2}q'\;.
\eea
These equations imply the variables 
${\cal{G}}^{\alpha\beta}$, 
$G^{\alpha\beta}$ and $\theta^{\alpha\beta}$ depend on the 
parameters $q$ and $q'$, which again are arbitrary. For branes with higher
dimensions, more free parameters appear in the open string variables and
in the propagator.

\end{document}